\documentclass[reprint,superscriptaddress,aps,twocolumn,showpacs]{revtex4-1}
        \usepackage[vcentermath]{youngtab}
\usepackage{amssymb}
\usepackage{amsmath}
\usepackage{graphicx}
\usepackage{multirow}
\usepackage{hyperref}
\usepackage{longtable}

\allowdisplaybreaks

\begin{document}

\title{Pentaquark components in low-lying baryon resonances}

\author{K. Xu}
\email[]{gxukai1123@gmail.com}
\author{A. Kaewsnod}
\author{Z. Zhao}
\affiliation{School of Physics and Center of Excellence in High Energy Physics and Astrophysics, Suranaree University of Technology, Nakhon Ratchasima 30000, Thailand}
\author{X. Y. Liu}
\affiliation{School of Physics and Center of Excellence in High Energy Physics and Astrophysics, Suranaree University of Technology, Nakhon Ratchasima 30000, Thailand}
\affiliation{School of Mathematics and Physics, Bohai University, Jinzhou 121013, China}
\author{S. Srisuphaphon}
\affiliation{Department of Physics, Faculty of Science, Burapha University, Chonburi 20131, Thailand}
\author{A. Limphirat}
\author{Y. Yan}
\email[]{yupeng@sut.ac.th}
\affiliation{School of Physics and Center of Excellence in High Energy Physics and Astrophysics, Suranaree University of Technology, Nakhon Ratchasima 30000, Thailand}

\date{\today}

\begin{abstract}
\indent  We study pentaquark states of both light $q^4\bar q$ and hidden heavy $q^3 Q\bar Q$ (q = u,d,s quark in SU(3) flavor symmetry; Q = c, b quark) systems with a general group theory approach in the constituent quark model, and the spectrum of light baryon resonances in the ansatz that the $l=1$ baryon states may consist of the $q^3$ as well as $q^4\bar q$ pentaquark component. The model is fitted to ground state baryons and light baryon resonances which are believed to be normal three-quark states. The work reveals that the $N(1535)1/2^{-}$ and $N(1520)3/2^-$ may consist of a large $q^4\bar q$ component while the $N(1895)1/2^{-}$ and $N(1875)3/2^-$ are respectively their partners, and the $N^+(1685)$ might be a $q^4\bar q$ state. By the way, a new set of color-spin-flavor-spatial wave function for $q^3 Q\bar Q$ systems in the compact pentaquark picture are constructed systematically for studying hidden charm pentaquark states.

\keywords{Permutation group, Spatial wave function, Mixing angle, Pentaquark mass spectra}
\end{abstract}

\maketitle

\section{Introduction}\label{sec:Int}
\indent  Baryon resonance spectrum has been studied over decades, but theoretical
results are still largely inconsistent with experimental data. Except for the ground state baryons, even the low-lying resonances, for example, the Roper resonance $N(1440)$, $N(1520)$ and $N(1535)$ have been of an ordering problem. Theoretical works in the three-quark picture always predict a larger mass for the lowest positive-parity state $N(1440)$ than for the lowest negative-parity states $N(1520)$ and $N(1535)$ \cite{Capstick2000}. Since the discoveries of $N(1895)1/2^{-}$, $N(1875)3/2^-$, $\Delta(1900)1/2^{-}$, and $\Delta(1940)3/2^-$ \cite{Aniso2012}, these states and other baryon resonance states near 1900 MeV have not been well explained in conventional constituent quark models \cite{Ron2015,Ron2018,Workman2012,Hunt2019}.

\indent By applying the new approaches of photoproduction and electroproduction experiments, more baryon resonances have been discovered and confirmed \cite{Aniso2012,Ron2015,Ron2018,Workman2012,Hunt2019,Sok2015} and the internal structures of some resonance states have been revealed by the properties including Breit-Wigner amplitudes, transitions amplitudes, and form factors \cite{Aniso2012,Segovia2015,Burkert2019,Olb2018,Ram2017}. The Roper electroproduction amplitudes \cite{Segovia2015} has proven us that it is mainly the nucleon first radial excitation as interpreted in the review paper \cite{Burkert2019}. That the decay width of $\Gamma_{N(1535)\to N \eta} \equiv (65\pm 25 \;\rm MeV)$ is as large as $\Gamma_{N(1535)\to N \pi} \equiv (67.5\pm 19\; \rm MeV)$ \cite{PDG} indicates that N(1535) may couple to the $\eta$ meson much more strongly than predicted by flavor symmetry \cite{Olb2018}. The strangeness component in $N(1535){1/2^-}$ is shown to account for the mass ordering of N(1440) and N(1535) \cite{Liu2006}, and it is claimed that $s\bar s$ pair contribution is important to the properties of the nucleon in Ref. \cite{Bijker2012}. As for the other lowest orbital excited state $N(1520){3/2^-}$, the branching ratio of $\Gamma_{\eta N} / \Gamma_{tot} $ is less than $1\%$ \cite{PDG} which reveals that there is little strange component contribution. It is also stated that $\gamma N \to N(1520)$ form factors are dominated by the meson cloud contributions which means N(1520) may not be pure $q^3$ state but include the extrinsic $q\bar q$ pair contribution in the form of $q^4 \bar q$ components \cite{Ram2017}. And the baryon states including pentaquark components have also been studied in the light quark sectors for Roper resonance~\cite{Li2006}, $N(1535)$~\cite{An2006,Jido2008,An2009} to give a better explanation of the experimental results like transitions amplitudes and form factors.

\indent In this work we study the role of pentaquark components in low-lying baryon resonance states. The constructions of light pentaquark wave functions in the Yamanouchi technique have been
formulated in the previous works~\cite{Sorakrai2012,Kai2014,Sorakrai2016,Kai2019PRC}. As a consequence, the light baryon resonance spectrum is newly reproduced by mixing three quark and pentaquark components. And we extend the group theory approach to hidden heavy pentaquarks in the SU(3) flavor symmetry, where the pentaquark wave functions for the $q^3 Q\bar Q$ systems are systematically constructed in the harmonic oscillator interaction and applied as complete bases to evaluate hidden charm and bottom pentaquark mass spectra for all possible quark configurations
and interactions of other types.

\indent  The paper is organized as follows: We briefly review in Sec.~\ref{sec:GTM} the
constituent quark model extensively described in our previous work \cite{Kai2019PRC}, and predetermined all the model parameters by comparing the theoretical and experimental masses of all the ground state baryons and low-lying $q^3$ baryon resonance states. The baryon masses in the $q^3$ picture are also presented in Sec.~\ref{sec:GTM}. In Sec.~\ref{sec:LQS} we derive the mass spectra of light $q^4 \bar q$ pentaquark states, and to reproduce the negative-parity nucleon and $\Delta$ resonances below 2 GeV by introducing light pentaquark components in three-quark baryon states. The wave functions of $q^3 Q\bar Q$ systems are constructed in the harmonic oscillator interaction for all possible quark configurations and applied as complete bases to evaluate hidden heavy pentaquark mass spectra in Sec.~\ref{sec:HID}. A summary is given in Sec.~\ref{sec:SUM}. The details of $q^3$ wave functions as well as the construction of $q^3 Q\bar Q$ pentaquark wave functions are shown in the Appendices.

\section{\label{sec:GTM}THEORETICAL MODEL}

\indent A group theory approach to construct the wave functions for baryon and pentaquark states has been described in Refs.~\cite{Sorakrai2012,Kai2014,Sorakrai2016,Kai2019PRC}, and we refer the readers to those works for details. Here, we just present the general Hamiltonian for multiquark systems,
\begin{flalign}\label{eqn::ham}
&H =H_0+ H_{hyp}^{OGE}, \nonumber \\
&H_{0} =\sum_{k=1}^{N} (m_k+\frac{p_k^2}{2m_{k}})+\sum_{i<j}^{N}(-\frac{3}{8}\lambda^{C}_{i}\cdot\lambda^{C}_{j})(A_{ij} r_{ij}-\frac{B_{ij}}{r_{ij}}),  \nonumber \\
&H_{hyp}^{OGE} = -C_{OGE}\sum_{i<j}\frac{\lambda^{C}_{i}\cdot\lambda^{C}_{j}}{m_{i}m_{j}}\,\vec\sigma_{i}\cdot\vec\sigma_{j},
\end{flalign}
where  $A_{ij}$ and $B_{ij}$ are mass-dependent coupling  parameters, taking the form
\begin{eqnarray}
A_{ij}= a \sqrt{\frac{m_{ij}}{m_u}},\;\;B_{ij}=b \sqrt{\frac{m_u}{m_{ij}}}.
\end{eqnarray}
with $m_{ij}$ being the  reduced mass of $i$th and $j$th  quarks, defined as $\;m_{ij}=\frac{2 m_i m_j}{m_i+m_j}$ which corresponds to the relative Jacobi coordinates of two-body system in Appendix \ref{sec:AP2}.
The hyperfine interaction, $H_{hyp}^{OGE}$ includes only one-gluon exchange contribution,
where $C_{OGE} = C_m\,m_u^2$, with $m_u$ being the constituent $u$ quark mass and $C_m$ a constant. $\lambda^C_{i}$ in the above equations are the generators of color SU(3) group.

\indent The model parameters are determined by fitting the theoretical results to the experimental data of the mass of all the ground state baryons, namely, eight light baryon isospin states, seven charm baryon states, and six bottom baryon states as well as light baryon resonances of energy level $N\leq2$, including the first radial excitation state $N(1440)$ with mass at 1.5 GeV and a number of orbital excited $l=1$ and $l=2$ baryons. All these baryons are believed to be mainly $3q$ states whose masses were taken from Particle Data Group~\cite{PDG}. The least squares method is applied to minimize the weighted squared distance $\delta^2$,
\begin{eqnarray}\label{eq:LSM}
\delta^2=\sum_{i=1}^N \omega_i \frac{(M^{exp}-M^{cal})^2}{{M^{exp}}^2}
\end{eqnarray}
where $\omega_i$ are weights being $1$ for all the states except for N(939) and $\Delta(1232)$ which are set to be 100, $M^{exp}$ and $M^{cal}$ are respectively the experimental and theoretical masses. Listed in Tables~\ref{tab:groundB}, \ref{tab:posN}, \ref{tab:negN}, and \ref{tab:delta} are the theoretical masses which are calculated in the Hamiltonian in Eq. (\ref{eqn::ham}) in the $q^3$ picture and fitted to the experimental data. Possible assignments of the theoretical results of excited nucleon and $\Delta$ resonances below 2.2 GeV to all the known baryon states are presented in Tables \ref{tab:posN}, \ref{tab:negN}, and \ref{tab:delta} following the $SU(6)_{SF}$ representations. The orbital-spin-flavor wave functions of $q^3$ baryon states are listed in Appendix \ref{sec:AP1}.
\begin{table}[b]
\caption{Ground state baryons applied to fit the model parameters. The last column shows the deviation between the experimental and theoretical mean values, $D=100\cdot (M^{exp}-M^{cal})/M^{exp}$. $M^{exp}$ are taken from PDG~\cite{PDG}.}
\label{tab:groundB}
\begin{ruledtabular}
\begin{tabular}{lccc}
 Baryon & $M^{exp}  {\rm (MeV)}$& $M^{cal} {\rm (MeV)}$  & D ($\%$)
\\
\hline
 N(939) & 939  &  939 & 0
\\
 $\Delta(1232)$ & 1232 & 1232 &  0
\\
 $\Lambda(1116)$ & 1116 & 1129 &  -1.16
\\
 $\Sigma(1193)$ & 1193  & 1163 &  2.56
\\
 $\Sigma^{*}(1385)$ & 1385 & 1372 & 0.97
\\
 $\Xi(1318)$ & 1318 & 1329 &  -0.83
\\
 $\Xi^{*}(1530)$ & 1533  & 1510 &  1.49
\\
 $\Omega(1672)$ & 1672 & 1662 &  0.62\\
\hline
 $\Lambda_C(2286)$ & 2286 &  2272 &  0.62
\\
 $\Sigma_C(2455)$ & 2454 & 2428 & 1.06
\\
 $\Sigma^{*}_C(2520)$ & 2518 & 2486 &  1.26
\\
 $\Xi_C(2470)$ & 2469 & 2489 &  -0.82
\\
 $\Xi^{*}_C(2645)$ & 2646 & 2633 &  0.47
\\
 $\Omega_C(2695)$ & 2695.& 2751 &  -2.07
\\
 $\Omega^{*}_C(2770)$ & 2766 & 2789 &  -0.84\\
\hline
 $\Lambda_B(5620)$ & 5620 &  5599 &  0.37
\\
 $\Sigma_B(5811)$ & 5811 & 5781 &  0.51
\\
 $\Sigma^{*}_B(5832)$ & 5832 & 5801 & 0.54
\\
 $\Xi_B(5792)$ & 5792 & 5819 &  -0.47
\\
 $\Xi^{*}_B(5945)$ & 5950 & 5953 &  -0.05
\\
 $\Omega_B(6046)$ & 6046 & 6097 &  -0.84\\
\end{tabular}
\end{ruledtabular}
\end{table}

The 3 model coupling constants and 4 constituent quark masses are fitted,
\begin{eqnarray}\label{eq:nmo}
&
m_u = m_d = 327 \ {\rm MeV}\,, \quad
m_s = 498 \ {\rm MeV}\,, \nonumber\\
&
m_c = 1642 \ {\rm MeV}\,, \quad
m_b = 4960 \ {\rm MeV}\,, \nonumber\\
&
C_m   =  18.3 \ {\rm MeV}, \quad
a     = 49500 \ {\rm MeV^2}, \quad
b     =  0.75 \nonumber\\
\end{eqnarray}
Similar model parameters were obtained in the previous work \cite{Kai2019PRC}. The parameters fixed in the work are slightly different from the preliminary ones since charm and bottom baryons are included and more accurate method is used for the model fixing. And the u and d constituent quark mass is closer to the quark mass, 330 ${\rm MeV}$ which was determined by the baryon magnetic moments \cite{Reju1975}.

\begin{table}[t]
\caption{Nucleon resonances of positive parity applied to fit the model parameters.}
\label{tab:posN}
\begin{ruledtabular}
\begin{tabular}{lcccc}
 $(\Gamma, {}^{2s+1} D,N,L^P)$  & Status & $J^{P}$ &  $M^{exp}  {\rm (MeV)}$ & $M^{cal} {\rm (MeV)}$
\\
\hline
 $N(56,{}^2 8,0,0^+)$ & **** & $\frac{1}{2}^{+}$ & 939 & 939
\\
 $N(56,{}^2 8,2,0^+)$ & **** & $\frac{1}{2}^{+}$ & N(1440) & 1499
\\
 $N(56,{}^2 8,2,2^+)$ & **** &$\frac{5}{2}^{+}$ & N(1720) & 1655
\\
 $N(56,{}^2 8,2,2^+)$ & **** & $\frac{3}{2}^{+}$ & N(1680) & 1655
\\
 $N(20,{}^2 1,2,1^+)$ & *** &$\frac{1}{2}^{+}$ & N(1880) & 1749
\\
 $N(20,{}^4 1,2,1^+)$ & - &$\frac{3}{2}^{+}$ & missing & 1749
\\
 $N(70,{}^2 10,2,0^+)$ & **** &$\frac{1}{2}^{+}$ & N(1710) & 1631
\\
 $N(70,{}^4 10,2,0^+)$ & ****  &$\frac{3}{2}^{+}$ & N(1900) & 1924
\\
 $N(70,{}^2 10,2,2^+)$ & - &$\frac{3}{2}^{+}$ & missing & 1702
\\
 $N(70,{}^2 10,2,2^+)$ & ** &$\frac{5}{2}^{+}$ & N(1860) & 1702
\\
 $N(70,{}^4 10,2,2^+)$ & ***  &$\frac{1}{2}^{+}$ & N(2100) & 1994
\\
 $N(70,{}^4 10,2,2^+)$ & *  &$\frac{3}{2}^{+}$ & N(2040) & 1994
\\
 $N(70,{}^4 10,2,2^+)$ & **  &$\frac{5}{2}^{+}$ & N(2000) & 1994
\\
 $N(70,{}^4 10,2,2^+)$ & **  &$\frac{7}{2}^{+}$ & N(1990) & 1994
\\
\end{tabular}
\end{ruledtabular}
\end{table}

\begin{table}[t]
\caption{Resonances of negative-parity applied to fit the model parameters.}
\label{tab:negN}
\begin{ruledtabular}
\begin{tabular}{lcccc}
 $(\Gamma,{}^{2s+1} D,N,L^P)$  & Status & $J^{P}$ & $M^{exp}  {\rm (MeV)}$ & $M^{cal} {\rm (MeV)}$
\\
\hline
 $N(70,{}^2 10,1,1^-)$ & **** &$\frac{3}{2}^{-}$ & N(1520) & 1380
\\
 $N(70,{}^2 10,1,1^-)$ & **** &$\frac{1}{2}^{-}$ & N(1535) & 1380
\\
 $N(70,{}^4 10,1,1^-)$ & **** &$\frac{1}{2}^{-}$ & N(1650) & 1672
\\
 $N(70,{}^4 10,1,1^-)$ & **** &$\frac{5}{2}^{-}$ & N(1675) & 1672
\\
 $N(70,{}^4 10,1,1^-)$ & *** &$\frac{3}{2}^{-}$ & N(1700) &1672
\\
  $\Delta(70,{}^2 10,1,1^-)$ & **** &$\frac{1}{2}^{-}$ & $\Delta$(1620) & 1380
\\
 $\Delta(70,{}^2 10,1,1^-)$ & **** &$\frac{3}{2}^{-}$ & $\Delta$(1700)  & 1380 \\
\end{tabular}
\end{ruledtabular}
\end{table}

\begin{table}
\caption{$\Delta$ resonance of positive parity applied to fit the model parameters.}
\label{tab:delta}
\begin{ruledtabular}
\begin{tabular}{lcccc}
 $(\Gamma,{}^{2s+1} D,N,L^P)$  & Status & $J^{P}$ &   $M^{exp}  {\rm (MeV)}$ & $M^{cal} {\rm (MeV)}$
\\
\hline
 $\Delta(56,{}^4 8,0,0^+)$ & **** & $\frac{3}{2}^{+}$ & $\Delta$(1232) & 1232
\\
 $\Delta(56,{}^4 8,2,0^+)$ & *** & $\frac{3}{2}^{+}$ & $\Delta$(1600) & 1791
\\
 $\Delta(56,{}^4 8,2,2^+)$ & **** &$\frac{5}{2}^{+}$ & $\Delta$(1905) & 1947
\\
 $\Delta(56,{}^4 8,2,2^+)$ & **** & $\frac{1}{2}^{+}$ & $\Delta$(1910) & 1947
\\
 $\Delta(56,{}^4 8,2,2^+)$ & *** & $\frac{3}{2}^{+}$ & $\Delta$(1920) & 1947
\\
 $\Delta(56,{}^4 8,2,2^+)$ & **** & $\frac{7}{2}^{+}$ & $\Delta$(1950) & 1947
\\
 $\Delta(70,{}^2 10,2,0^+)$ & * & $\frac{1}{2}^{+}$ &  $\Delta$(1750) & 1631
\\
 $\Delta(70,{}^2 10,2,2^+)$ & - & $\frac{3}{2}^{+}$ &  $missing$ & 1702
\\
 $\Delta(70,{}^2 10,2,2^+)$ & - & $\frac{5}{2}^{+}$ &  $missing$ & 1702
\\
\end{tabular}
\end{ruledtabular}
\end{table}
In general, all the ground state baryons are well described, with the maximum deviation less than $3\%$. For excited baryon states, the Roper resonance as the first radial excited state gets a mass around 1.5 GeV which does not agree well with the pole mass on PDG \cite{PDG}, but has a 0.56 GeV gap between the ground state nucleon, close to the gap 0.55 GeV between the two lowest-magnitude $J^P=1/2^+$ poles in Refs. \cite{Segovia2015,Burkert2019}. The lowest negative-parity nucleon states turn out to be lower than the Roper resonance just as other predictions of the conventional constituent quark models. We assume that the lowest negative-parity baryon resonances may consist of the $q^3$ component as well as the $q^4\,\bar q$ pentaquark component. The spin-orbit interactions are not included in this work, so the states in the same spatial-spin-flavor configuration as shown in Appendix \ref{sec:AP1} have
the same mass value. Except for the two missing $\Delta(70,{}^2 10,2,2^+)$ states and the two missing nucleon states $N(20,{}^2 1,2,1^+)$ and $N(70,{}^2 10,2,2^+)$, most positive-parity states are reasonably reproduced.

\section{LIGHT QUARK SPECTRUM}\label{sec:LQS}
\subsection{Mass of $q^4\bar q$ pentaquark states}
The mass spectra of the ground state $q^4\bar q$ and $q^3s\bar s$ pentaquarks are evaluated in the Hamiltonian in Eq. (\ref{eqn::ham}), by applying the complete bases of the pentaquark wave functions derived in our previous work \cite{Kai2019PRC}. Listed in Tables \ref{tab:d1} and \ref{tab:d2} are the theoretical results, with the model parameters fixed in the previous section. Comparing to other works \cite{Bijker2004,Huang2018} for $q^4 \bar q$ and $q^3 s\bar s$ hidden strange pentaquark states, the model here employs much less model parameters and predict relatively higher mass spectra.
It is predicted in the calculation that the pentaquark state with the ${[31]_{FS}[22]_{F}[31]_{S}}$ configuration and the quantum numbers $I(J^P)=\frac1{2}(\frac1{2}^{-})$
has the lowest mass, 1683 ${\;\rm MeV}$ which is quite close to the mass of the isospin-$1/2$ narrow resonance $N^{+}(1685)$. One may make a bold guess that this $N^{+}(1685)$ resonance could be the lowest pentaquark state.
\begin{table}[t]
\caption{\label{tab:d1}$q^4\bar q$ ground state pentaquark masses.}
\begin{ruledtabular}
\begin{tabular}{lcc}
 $q^4\bar q$ configurations & $J^{P}$ &  $M(q^{4}\bar q)$  ({\rm MeV})
\\
\hline
 $\Psi^{csf}_{[211]_{C}[31]_{FS}[4]_{F}[31]_{S}}(q^{4}\bar q)$ & $\frac{1}{2}^{-}$, $\frac{3}{2}^{-}$
 & 2562, 2269
\\
 $\Psi^{csf}_{[211]_{C}[31]_{FS}[31]_{F}[4]_{S}}(q^{4}\bar q)$ & $\frac{3}{2}^{-}$, $\frac{5}{2}^{-}$
 & 2025, 2269
\\
 $\Psi^{csf}_{[211]_{C}[31]_{FS}[31]_{F}[31]_{S}}(q^{4}\bar q)$ & $\frac{1}{2}^{-}$,
$\frac{3}{2}^{-}$   & 2123, 2049
\\
 $\Psi^{csf}_{[211]_{C}[31]_{FS}[31]_{F}[22]_{S}}(q^{4}\bar q)$ & $\frac{1}{2}^{-}$   & 2025
\\
 $\Psi^{csf}_{[211]_{C}[31]_{FS}[22]_{F}[31]_{S}}(q^{4}\bar q)$ & $\frac{1}{2}^{-}$,
$\frac{3}{2}^{-}$  & 1683, 2049 \\
\end{tabular}
\end{ruledtabular}
\end{table}

\begin{table}[b]
\caption{$q^3 s\bar s$ ground state pentaquark masses.}
\label{tab:d2}
\begin{ruledtabular}
\begin{tabular}{lcc}
 $q^4\bar q$ configurations & $J^{P}$  &  $M(q^3 s\bar s)$  ({\rm MeV})
\\
\hline
 $\Psi^{csf}_{[211]_{C}[31]_{FS}[4]_{F}[31]_{S}}(q^3s\bar s)$ & $\frac{1}{2}^{-}$, $\frac{3}{2}^{-}$
&  2762, 2586
\\
 $\Psi^{csf}_{[211]_{C}[31]_{FS}[31]_{F}[4]_{S}}(q^3s\bar s)$ & $\frac{3}{2}^{-}$, $\frac{5}{2}^{-}$
 & 2420, 2546
\\
 $\Psi^{csf}_{[211]_{C}[31]_{FS}[31]_{F}[31]_{S}}(q^3s\bar s)$ & $\frac{1}{2}^{-}$,
$\frac{3}{2}^{-}$   & 2448, 2414
\\
 $\Psi^{csf}_{[211]_{C}[31]_{FS}[31]_{F}[22]_{S}}(q^3s\bar s)$ & $\frac{1}{2}^{-}$  & 2393
\\
 $\Psi^{csf}_{[211]_{C}[31]_{FS}[211]_{F}[31]_{S}}(q^3s\bar s)$ & $\frac{1}{2}^{-}$,
$\frac{3}{2}^{-}$   & 2032, 2243
\\
 $\Psi^{csf}_{[211]_{C}[31]_{FS}[211]_{F}[22]_{S}}(q^3s\bar s)$ & $\frac{1}{2}^{-}$   &
2165
\\
 $\Psi^{csf}_{[211]_{C}[31]_{FS}[22]_{F}[31]_{S}}(q^3s\bar s)$ & $\frac{1}{2}^{-}$,
$\frac{3}{2}^{-}$  & 2135, 2354\\
\end{tabular}
\end{ruledtabular}
\end{table}

\subsection{Possible mixtures of $q^3$ and $q^4\,\bar q$ states}
\begin{table}[b]
\caption{The mixture of $q^3$ and $q^4\,\bar q$ components. All four $q^3$ states take the same mass, $1380 \;{\rm MeV}$. The chosen pentaquark states and masses are listed as $q^4 \bar q$ configuration and $q^4 \bar q$ Mass (in MeV) from Tables \ref{tab:d1} and \ref{tab:d2}.}
\label{fall}
\begin{ruledtabular}
\begin{tabular}{lccccc}
 $\psi_1$ State &  $J^{P}$ & $\theta$ & $\psi_2$ State & $q^4 \bar q$ configuration & $q^4 \bar q$ Mass \\
\hline
1530 & $\frac{1}{2} ^{-}$ & $i35.2^{\circ}$ &  1882 & $q^{3} s\bar s_{[211]_{F}[31]_{S}}$ & 2032\\

 1515 & $\frac{3}{2} ^{-}$ & $i32.6^{\circ}$ & 1899 & $q^{4}\bar q_{[31]_{F}[4]_{S}}$ & 2025\\
 
  $\phantom{-}$ & $\phantom{-}$ & $i31.7^{\circ}$ & 1914 & $q^{4}\bar q_{[22]_{F}[31]_{S}}$ & 2049\\

  1610 & $\frac{1}{2} ^{-}$ & $i46.4^{\circ}$ & 1893 & $q^{4} \bar q_{[31]_{F}[31]_{S}}$ & 2123   \\

 1710 & $\frac{3}{2} ^{-}$ & $i51.5^{\circ}$ & 2024 & $q^{3} s\bar s_{[22]_{F}[31]_{S}}$ & 2354 \\
\end{tabular}
\end{ruledtabular}
\end{table}

\indent Ground state pentaquarks always have a negative parity, thus only $l=1$ nucleon and $\Delta$ orbitally excited states could mix with ground state pentaquarks. Considering the low theoretical masses for the $N(1535)$ and $N(1520)$ resonances in the $q^3$ picture and their quantum numbers, it is natural to assume that the two baryon resonances may include both the $q^3$ and $q^4\,\bar q$ pentaquark component contributions. The wave function of these baryon resonances may be expressed as linear combinations of the $q^3$ state and 
$q^4\,\bar q$ pentaquark states which have the same quantum numbers as the $q^3$ state,
\begin{eqnarray}\label{eq:mix}
a_0 |q^3\rangle +\sum_{\alpha} a_\alpha |q^4\bar q\rangle^\alpha\,. 
\end{eqnarray}
In principle, one can determine the coefficients $a_\alpha$ by solving the coupled equations of all channels including not only the coupling between the $q^3$ and $q^4\,\bar q$ states and the coupling between the $q^4\,\bar q$ states, but also the contributions of hidden channels such as meson-baryon ones. The mass matrix is usually not Hermitian but complex, thus the bare states and physical states cannot be linked by an unitary transformation. In this work we simplify the problem to the simplest case that the $q^3$ state mixes with only one $q^4\,\bar q$ pentaquark state which has the lowest mass, eliminating other pentaquark states and meson-baryon channels. As a result, the $2\times 2$ mass matrix will be highly complex, which may be eigendiagonalized by the transformation,
\begin{eqnarray}\label{eq:mixtheta}
\psi_1 &=& \cos \theta |q^3\rangle -\sin \theta |q^4\bar q\rangle \,, \nonumber\\
\psi_2 &=& \sin \theta |q^3\rangle + \cos \theta |q^4\bar q\rangle.
\end{eqnarray}
where $\psi_1$ and $\psi_2$ are respectively the lower and higher  negative-parity physical states, and the mixing angle $\theta$ between the $q^3$ and $q^4 \bar q$ states is generally complex. The masses of the physical states, $M_{\psi_1}$ and $M_{\psi_2}$ are derived as follows:
\begin{eqnarray}\label{eq:sol}
M_{\psi_1} &=& M_{q^3} \cos^2 \theta+M_{q^4\bar q}\sin^2 \theta -m_\delta \,, \nonumber\\
M_{\psi_2} &=& M_{q^3} \sin^2 \theta+M_{q^4\bar q}\cos^2 \theta +m_\delta  \,, \nonumber\\
m_\delta &=& \frac{ (M_{q^4\bar q}-M_{q^3})}{2}\tan 2\theta \sin 2\theta
\end{eqnarray}
The mixing angle $\theta$ in Eq. (\ref{eq:sol}) is determined by adjusting the lower negative-parity states $\psi_1$ to $N(1535)$, $N(1520)$, $\Delta(1620)$, and $\Delta(1700)$. With both the real and imaginary part of the mixing angle in the domain of $(0, \pi/2)$, the mixing angle and the $M_{\psi_2}$ can be determined without duplication from Eq. (\ref{eq:sol}). Thus, one gets four pairs of
mixed states as shown in Table \ref{fall} with all $Re(\theta)=0$. $N(1520)3/2^-$ and $N(1875)3/2^-$ form a nonstrange pair, and the $N(1535)1/2^{-}$ and $N(1895)1/2^{-}$ form a strange pair for the nucleon resonances while the $\Delta(1620)1/2^{-}$ and $\Delta(1900)1/2^{-}$ form a nonstrange pair, and the $\Delta(1700)3/2^-$ and $\Delta(1940)3/2^-$ form a strange pair for the $\Delta$ resonances. For the pair of N(1520) and N(1875), we have shown in Table \ref{fall} the results with both the pentaquark states $q^{4}\bar q_{[31]_{F}[4]_{S}}$ (2025 MeV) and  $q^{4}\bar q_{[22]_{F}[31]_{S}}$ (2049 MeV) mixed with the $q^3$ state. In the present model one can not rule out either of them.


The mass spectrum of the negative-parity nucleon and $\Delta$ resonances are listed in Table \ref{tab:negpanu} in the $q^3$ and $q^4 \bar q$ picture. $N(1650)$, $N(1675)$, and $N(1700)$ are assumed to be mainly pure $q^3$ states since the $q^3$ picture reproduces their masses well, as shown in Table \ref{tab:negN}, and hence there is no mixing with pentaquark states. $N(1685)$ could be the lowest pure pentaquark state. The others are $q^3$ and $q^4\,\bar q$ mixing states taken from Table \ref{fall}.

\begin{table}[t]
\caption{Masses of negative-parity resonances after including ground state pentaquark components. The theoretical masses of $N(1535)$, $N(1520)$, $\Delta(1620)$, and $\Delta(1700)$ states take the mean values of their Breit-Wigner mass from \cite{PDG}.}\label{tab:negpanu}
\begin{ruledtabular}
\begin{tabular}{lcccc}
 Resonance  & Status & $J^{P}$ & $M^{exp}$ ${\rm (MeV)}$ & $M^{cal}$ ${\rm (MeV)}$
\\
\hline
 $N(1520)$ & **** &$\frac{3}{2}^{-}$ & 1510-1520 & 1515
\\
 $N(1535)$ & **** &$\frac{1}{2}^{-}$ & 1515-1545 & 1530
\\
 $N(1650)$ & **** &$\frac{1}{2}^{-}$ & 1645-1670 &1672
\\
 $N(1675)$ & **** &$\frac{5}{2}^{-}$ & 1670-1680 &1672
\\
 $N(1685)$ & * &$\frac{1}{2}^{-} ?$ & 1665-1675 & 1683
\\
 $N(1700)$ & *** &$\frac{3}{2}^{-}$ & 1650-1750 &1672
\\
 $N(1875)$ & *** &$\frac{3}{2}^{-}$ & 1850-1920 &1899/1914
\\
 $N(1895)$ & **** &$\frac{1}{2}^{-}$ & 1870-1920 &1882
\\
 $\Delta(1620)$ & **** &$\frac{1}{2}^{-}$ & 1590-1630 &1610
\\
 $\Delta(1700)$ & **** &$\frac{3}{2}^{-}$ & 1690-1730 &1710
\\
 $\Delta(1900)$ & *** &$\frac{1}{2}^{-}$ & 1840-1920 &1893
\\
 $\Delta(1940)$ & ** &$\frac{3}{2}^{-}$ & 1940-2060 & 2024
\\
\end{tabular}
\end{ruledtabular}
\end{table}
The conventional constituent quark models have failed to describe the higher nucleon and $\Delta$ resonance states near 1900 MeV \cite{Ron2015,Ron2018,Workman2012,Hunt2019}. In this constituent quark model with a color dependent Cornell-like potential, however, we have given not only the possible theoretical interpretations for $N(1895)1/2^{-}$, $N(1875)3/2^-$, $\Delta(1900)1/2^{-}$, and $\Delta(1940)3/2^-$ states as negative-parity partners of the well known nucleon and $\Delta$ resonances, but also effectively solved the long-standing ordering problem of $N(1440)$, $N(1520)$ and $N(1535)$ by mixing the $q^3$ and $q^4\,\bar q$ components. 

In general, a $q^3$ state may mix with two or even more $q^4\,\bar q$ states as well as meson-baryon ones. However, the present work can not give more meaningful information by including two or more pentaquark states in the mixture. A better understanding of $N(1440)$, $N(1520)$ and $N(1535)$ may be achieved  by studying the helicity amplitude of $N(1440)$, $N(1520)$ and $N(1535)$ with both $q^3$ and $q^4\,\bar q$ state contributions since there are much more sensitive experimental data available.

\section{$q^3 Q\bar Q$ PENTAQUARK SPECTRUM}\label{sec:HID}

\indent Motivated by the hidden-charm pentaquark candidates recently found by the LHCb Collaboration~\cite{LHCb3} we also calculate the mass spectra of hidden heavy pentaquarks of $q^3 Q\bar Q$ systems. The quark configurations and wave functions of the $q^3 Q\bar Q$ systems are derived in Appendix \ref{sec:AP2}. The spatial wave functions, which are derived in the harmonic oscillator quark-quark interaction and grouped in Appendix \ref{sec:AP2} according to the permutation symmetry, are employed as complete bases to study the $q^3 Q\bar Q$ systems described with the color dependent Hamiltonian in Eq. (\ref{eqn::ham}). 
The mass spectra of the hidden charm and hidden bottom pentaquarks of the $q^3$ color octet configuration are presented in Tables \ref{tab:d3} and \ref{tab:d4} separately.

\begin{table}[t]
\caption{\label{tab:d3}Ground hidden-charm pentaquark $q^3c\bar c$ mass spectrum, where the $q^3$ and $Q\bar Q$ components are in the color octet states.}
\begin{ruledtabular}
\begin{tabular}{lcc}
\multicolumn{1}{l}{\phantom{-}$q^3Q\bar Q$ configurations}&
   \multicolumn{1}{c}{$\phantom{-}$$J^{P}$}&
   \multicolumn{1}{c}{$M(q^{3} c\bar c)$(MeV)} \\[2pt]
\hline\\[-13pt]
  $\Psi^{csf}_{[21]_{C}[21]_{FS}[21]_{F}[21]_{S}}(q^3c\bar c)$ & $\frac{1}{2}^{-}$, $\frac{3}{2}^{-}$  & 4483, 4495\\[2pt]
  $\Psi^{csf}_{[21]_{C}[21]_{FS}[3]_{F}[21]_{S}}(q^3c\bar c)$ & $\frac{1}{2}^{-}$, $\frac{3}{2}^{-}$  & 4702, 4701\\[2pt]
  $\Psi^{csf}_{[21]_{C}[21]_{FS}[21]_{F}[3]_{S}}(q^3c\bar c)$ & $\frac{3}{2}^{-}$, $\frac{5}{2}^{-}$  & 4556, 4598\\
\end{tabular}
\end{ruledtabular}
\end{table}

\begin{table}[b]
\caption{\label{tab:d4}Ground hidden-bottom pentaquark $q^3 b\bar b$ mass spectrum, where the $q^3$ and $Q\bar Q$ components are in the color octet states.}
\begin{ruledtabular}
\begin{tabular}{lcc}
\multicolumn{1}{l}{\phantom{-}$q^3Q\bar Q$ configurations}&
   \multicolumn{1}{c}{$\phantom{-}$$J^{P}$}&
   \multicolumn{1}{c}{$M(q^{3} b\bar b)$(MeV)} \\[2pt]
\hline\\[-13pt]
  $\Psi^{csf}_{[21]_{C}[21]_{FS}[21]_{F}[21]_{S}}(q^3b\bar b)$ & $\frac{1}{2}^{-}$, $\frac{3}{2}^{-}$ & 10964, 10968 \\[2pt]
  $\Psi^{csf}_{[21]_{C}[21]_{FS}[3]_{F}[21]_{S}}(q^3b\bar b)$ & $\frac{1}{2}^{-}$, $\frac{3}{2}^{-}$ & 11183, 11183 \\[2pt]
  $\Psi^{csf}_{[21]_{C}[21]_{FS}[21]_{F}[3]_{S}}(q^3b\bar b)$ & $\frac{3}{2}^{-}$, $\frac{5}{2}^{-}$ & 11037, 11051 \\
\end{tabular}
\end{ruledtabular}
\end{table}

It's noted that the hidden-charm pentaquark mass spectra in this work is slightly higher than the three narrow pentaquarklike states, $P_c(4312)^{+}$, $P_c(4440)^{+}$, and $P_c(4457)^{+}$ measured by LHCb. The predicted values of 4483 and 4495 MeV for the lowest hidden-charm pentaquark in the ${[21]_{C}[21]_{FS}[21]_{F}[21]_{S}}$ configuration are close to the experimental values of 4440 and 4457 MeV, but still about 100-200 MeV higher than the $P_c(4312)^{+}$ state. The higher predicted $P_c$ masses may result from the compact spacial configuration in our pentaquark picture. The observed $P_c$ may probably be baryon-meson molecular states or mixtures of compact pentaqark states and molecules.   
For the hidden-bottom pentaquarks, the work predicts the mass of the ground states to be 10.9-11.2 GeV, lying below the threshold of a single bottom baryon and $B(B^*)$ mesons, which is consistent with other work \cite{Thomas2019}.

The newly observed $P_c$ states by the LHCb collaboration have been largely interpreted as hadronic molecule states since there are abundant charmed meson and charmed baryon thresholds available~\cite{LHCb3}. Within the molecular scenario, the mass spectrum \cite{Zhu2019,Chen2019,Liu2019,He2019,Oller2019,Xiao22019,Meng2019,Wu2019,Guo2019,Xiao12019,Voloshin2019,Sakai2019,Lin2019,Thomas2019} and dynamical properties~\cite{Guo2019,Xiao12019,Voloshin2019,Sakai2019,Lin2019,Thomas2019} have been successfully explained in various methods. The compact pentaquark interpretation works well \cite{Zhur2019,Giron2019,Cheng2019} when the parameters are fixed to both baryons and mesons. With the limited experimental results, the nature of $P_c$ states will keep as an open question in the near future.  

\section{Summary}\label{sec:SUM}

\indent The masses of low-lying $q^3$ states and ground $q^4\bar q$ states are evaluated, where all model parameters are predetermined by fitting the theoretical masses to the experimental data for the baryons which are believed to be mainly $3q$ states. In the work we have assumed that the Roper resonance is the first radial excitation state of nucleon.

It is interesting that the theoretical work predicts 
the pentaquark state with the ${[31]_{FS}[22]_{F}[31]_{S}}$ configuration and the quantum numbers $I(J^P)=\frac1{2}(\frac1{2}^{-})$ has the lowest mass, about 1680 MeV. One may make a bold guess that this $q^{4}\bar q$ pentaquark state could be the isospin-$1/2$ narrow resonance $N^{+}(1685)$ which can not be accommodated as a $q^3$ particle.

The work shows that the ordering problem of the $N(1440)$, $N(1520)$ and $N(1535)$ may be solved by introducing the $q^4\,\bar q$ contribution. The same calculation leads to that the $N(1895)1/2^{-}$, $N(1875)3/2^-$, $\Delta(1900)1/2^{-}$, and $\Delta(1940)3/2^-$ resonances may pair respectively with the $N(1535)1/2^{-}$, $N(1520)3/2^-$, $\Delta(1620)1/2^{-}$, and $\Delta(1700)3/2^-$ in the $q^3$ and $q^{4}\bar q$ interpretation. 

The mass spectra of ground hidden heavy pentaquark states $q^{3} Q\bar Q$ are accurately evaluated using the same predetermined model parameters. It is found that the hidden charm pentaquark states with the $[21]_{C}[21]_{FS}[21]_{F}[21]_{S}$ configuration have the lowest masses which are slightly larger than the LHCb results. In this communication, however, the work can not draw any conclusion about the nature of $P_c$ states.

\section{ACKNOWLEDGMENTS}
\indent  We are grateful for insightful comments and suggestions from Prof. Thomas Gutsche. This work is supported by Suranaree University of Technology (SUT) and the Office of the Higher Education Commission under the National Research University (NRU) project of Thailand. K.X. and Y.Y. acknowledge support from  SUT under Grant No. SUT-PhD/13/2554. S.S. acknowledges support from the Faculty of Science, Burapha University. A.K., Z.Z., and A.L. acknowledge support from SUT. X.Y.L. acknowledges support from Young Science Foundation from the Education Department of Liaoning Province, China (Project No. LQ2019009).

\appendix
%
%

\section{Explicit $q^3$ wave functions}\label{sec:AP1}
In this Appendix the $q^3$ color-orbital-spin-flavor wave functions with the principle quantum number $N\leq 2$ are listed in Table \ref{tab:osf}, where $\chi_i$, $\Phi_j$, and $\phi^{N'}_{L'M'y}$ are the spin, flavor, and spatial wave functions, respectively. The $SU(3)_F$ singlet states are excluded since only nucleon and $\Delta$ resonances are discussed. 

\begin{table*}[t]
\caption{\label{tab:osf}Explicit $q^3$ color-orbital-spin-flavor wave functions.}
\begin{ruledtabular}
\begin{tabular}{@{}lcccc}
\multicolumn{1}{l}{\phantom{-}}&
   \multicolumn{1}{c}{$\phantom{-}$$SU(6)_{SF}$}&
   \multicolumn{1}{c}{$l^P$}&
   \multicolumn{2}{c}{$SU(6)_{SF}$$\times O(3)$ wave functions} \\
   \hline
   \multicolumn{1}{l}{$\phantom{-}N$} &
   \multicolumn{1}{c}{$\phantom{-}$Representations} &
   \multicolumn{1}{c}{O(3)}&
   \multicolumn{1}{c}{$\phantom{-}$$SU(3)_F$ octet} &
   \multicolumn{1}{c}{$\phantom{-}$$SU(3)_F$ decuplet}\\[2pt]
\hline\\[-13pt]
   $\phantom{-}0$ & $\phantom{-}56$ & $0^{+}$ & $J^{P}=\frac{1}{2}^{+}$ & $J^{P}=\frac{3}{2}^{+}$\\[2pt]
    $\phantom{-}$ & $\phantom{-}$ & $\phantom{-}$ & $\frac{1}{\sqrt{2}}\psi^{c}_{[111]}\phi^0_{00s} (\Phi_{\lambda} \chi_{\rho}+\Phi_{\rho} \chi_{\lambda})$  & $\psi^{c}_{[111]}\phi^0_{00S}\Phi_{S}  \chi_{S}$\\[2pt]
    \hline
   $\phantom{-}1$ & $\phantom{-}70$ & $1^{-}$  & $ \; J^{P}=\frac{1}{2}^{-}\; , \;\frac{3}{2}^{-}$ & $J^{P}=\frac{1}{2}^{-}\; , \;\frac{3}{2}^{-}$\\[2pt]
   $\phantom{-}$ & $\phantom{-}$ & $\phantom{-}$  & $\frac{1}{2} \psi^{c}_{[111]}[\phi^1_{1m\rho} (\Phi_{\lambda} \chi_{\rho}+\Phi_{\rho} \chi_{\lambda})+\phi^1_{1m\lambda} (\Phi_{\rho} \chi_{\rho}-\Phi_{\lambda} \chi_{\lambda})]$  & $\frac{1}{\sqrt{2}}\psi^{c}_{[111]}\Phi_{S} (\phi^1_{1m\lambda} \chi_{\lambda}+\phi^1_{1m\rho} \chi_{\rho})$\\[2pt]
   $\phantom{-}$ & $\phantom{-}$ & $\phantom{-}$  & $J^{P}=\frac{1}{2}^{-}\; , \;\frac{3}{2}^{-},\;\frac{5}{2}^{-}$  & $\phantom{-}$\\[2pt]
   $\phantom{-}$ & $\phantom{-}$ & $\phantom{-}$ &  $\frac{1}{\sqrt{2}} \psi^{c}_{[111]}\chi_{S} (\phi^1_{1m\lambda} \Phi_{\lambda}+\phi^1_{1m\rho} \Phi_{\rho})$  & $\phantom{-}$\\[2pt]
   \hline
   $\phantom{-}2$ & $\phantom{-}56$ & $0^{+}$ & $J^{P}=\frac{1}{2}^{+}$ & $J^{P}=\frac{3}{2}^{+}$\\[2pt]
    $\phantom{-}$ & $\phantom{-}$ & $\phantom{-}$ & $\frac{1}{\sqrt{2}}\psi^{c}_{[111]}\phi^2_{00s} (\Phi_{\lambda} \chi_{\rho}+\Phi_{\rho} \chi_{\lambda})$  & $\psi^{c}_{[111]}\Phi_{S} \phi^2_{00S} \chi_{S}$\\[2pt]
   \hline
   $\phantom{-}$ & $\phantom{-}70$ & $0^{+}$ & $ J^{P}=\frac{1}{2}^{+}$ & $J^{P}=\frac{1}{2}^{+}$\\[2pt]
   $\phantom{-}$ & $\phantom{-}$ & $\phantom{-}$ & $\frac{1}{\sqrt{2}}\psi^{c}_{[111]}[\phi^2_{00\rho} (\Phi_{\lambda} \chi_{\rho}+\Phi_{\rho} \chi_{\lambda})+\phi^2_{00\lambda} (\Phi_{\rho} \chi_{\rho}-\Phi_{\lambda} \chi_{\lambda})]$  & $\frac{1}{2}\psi^{c}_{[111]}\Phi_{S} (\phi^2_{00\lambda} \chi_{\lambda}+\phi^2_{00\rho} \chi_{\rho})$\\[2pt]
   $\phantom{-}$ & $\phantom{-}$ & $\phantom{-}$ & $J^{P}=\frac{3}{2}^{+}$ & $\phantom{-}$\\[2pt]
   $\phantom{-}$ & $\phantom{-}$ & $\phantom{-}$ & $\frac{1}{\sqrt{2}} \psi^{c}_{[111]}\chi_{S} (\phi^2_{00\lambda} \Phi_{\lambda}+\phi^2_{00\rho} \Phi_{\rho})$  & $\phantom{-}$\\[2pt]
   \hline
      $\phantom{-}2$ & $\phantom{-}20$ & $1^{+}$  & $J^{P}=\frac{1}{2}^{+}\; , \;\frac{3}{2}^{+}$ & $\phantom{-}$\\[2pt]
   $\phantom{-}$ & $\phantom{-}$ & $\phantom{-}$ & $\psi^{c}_{[111]}\phi^2_{1mA} (\Phi_{\rho} \chi_{\rho}-\Phi_{\lambda} \chi_{\lambda})$  & $\phantom{-}$\\[2pt]
   \hline
    $\phantom{-}2$ & $\phantom{-}56$ & $2^{+}$  & $J^{P}=\frac{3}{2}^{+}\; , \;\frac{5}{2}^{+}$ & $J^{P}=\frac{1}{2}^{+}\; , \;\frac{3}{2}^{+} \;,\;\frac{5}{2}^{+} \;,\;\frac{7}{2}^{+}$\\[2pt]
    $\phantom{-}$ & $\phantom{-}$ & $\phantom{-}$ & $\frac{1}{\sqrt{2}} \psi^{c}_{[111]}\phi^2_{2mS} (\Phi_{\rho} \chi_{\rho}+\Phi_{\lambda} \chi_{\lambda})$  & $\psi^{c}_{[111]}\phi^2_{2mS} \Phi_{S} \chi_{S}$\\[2pt]
   \hline
   $\phantom{-}$ & $\phantom{-}70$ & $2^{+}$ & $ J^{P}=\frac{3}{2}^{+}\; , \;\frac{5}{2}^{+}$ & $J^{P}=\frac{3}{2}^{+}\; , \;\frac{5}{2}^{+}$\\[2pt]
   $\phantom{-}$ & $\phantom{-}$ & $\phantom{-}$ & $\frac{1}{2} \psi^{c}_{[111]}[\phi^2_{2m\rho} (\Phi_{\lambda} \chi_{\rho}+\Phi_{\rho} \chi_{\lambda})+\phi^2_{2m\lambda} (\Phi_{\rho} \chi_{\rho}-\Phi_{\lambda} \chi_{\lambda})]$  & $\frac{1}{\sqrt{2}}\psi^{c}_{[111]}\Phi_{S} (\phi^2_{2m\lambda} \chi_{\lambda}+\phi^2_{2m\rho} \chi_{\rho})$\\[2pt]
   $\phantom{-}$ & $\phantom{-}$ & $\phantom{-}$ & $J^{P}=\frac{1}{2}^{+}\; , \;\frac{3}{2}^{+} \;,\;\frac{5}{2}^{+} \;,\;\frac{7}{2}^{+}$ & $\phantom{-}$\\[2pt]
   $\phantom{-}$ & $\phantom{-}$ & $\phantom{-}$ & $\frac{1}{\sqrt{2}} \psi^{c}_{[111]}\chi_{S} (\phi^2_{2m\lambda} \Phi_{\lambda}+\phi^2_{2m\rho} \Phi_{\rho})$  & $\phantom{-}$\\
\end{tabular}
\end{ruledtabular}
\end{table*}

\section{Construction of pentaquark wave functions for $q^3 Q\bar Q$ system}\label{sec:AP2}

The construction of the $q^3 Q\bar Q$ pentaquark state follows the rule that $q^3 Q\bar Q$ state must be a color singlet and the $q^3 Q\bar Q$ wave function should be
antisymmetric under any permutation between identical quarks. Requiring the $q^3 Q\bar Q$ pentaquark to be a color singlet demands that the color part of the $q^3$ and $Q\bar Q$ must form a $[222]_1$ singlet state, there are two possible color configurations: the color part of the $q^3$ is a [111] singlet and the $Q\bar Q$ is also a singlet and the color part of the $q^3$ is a [21] octet and $Q\bar Q$ is also an octet.
The pentaquark state in the $q^3 Q\bar Q$ system with the $q^3$ color singlet configuration corresponds to the hadronic molecular pentaquark state which is not confined in our Hamiltonian. And the $q^3 Q\bar Q$ system in the compact pentaquark picture takes the $q^3$ color octet configuration. Requiring the wave function of the three-quark configuration to be antisymmetric, the spatial-spin-flavor part of $q^3$ is required to be $[21]$ state by conjugation, and directly couples with the spatial-spin-flavor part of $Q\bar Q$. First we study the total antisymmetric wave function for the $q^3$ color octet configuration,
\begin{eqnarray}\label{eqn::3q1}
\psi_{[3]_A} = \frac{1}{\sqrt{2}}
\left(\psi^{c}_{[21]_\lambda} \psi^{osf}_{[21]_\rho} -
\psi^{c}_{[21]_\rho} \psi^{osf}_{[21]_\lambda} \right)
\end{eqnarray}
with
\begin{eqnarray}\label{eqn::3q2}
\psi^{osf}_{[21]_{\rho,\lambda}} &=& \sum_{i,j=S,\rho,\lambda}b_{ij}\psi^{o}_{[X]_{i}}\psi^{sf}_{[Y]_{j}}, \nonumber \\
\psi^{sf}_{[Y]} &=& \sum_{i,j=S,\rho,\lambda}c_{ij}\psi^{s}_{[x]_{i}}\psi^{f}_{[y]_{j}},\nonumber \\
\psi^{s}_{[X]_{i}} &=& \{\psi^{s}_{[3]_S},\psi^{s}_{[21]_{\rho,\lambda}}\}, \nonumber \\
\psi^{f}_{[Y]_{j}} &=& \{\psi^{f}_{[3]_S},\psi^{f}_{[111]_A},\psi^{f}_{[21]_{\rho,\lambda}}\}
\end{eqnarray}
The total color wave function for $q^3 Q\bar Q$ pentaquark state takes the form,
\begin{eqnarray}\label{eqn::color}
 \Psi^{c}_{[21]_{j=\rho,\lambda}}&=&\frac{1}{\sqrt{8}} \sum^8_{i} \psi^{c}_{[21]^i_{j}}(q^3) \psi^{c}_{[21]^i_{j}}(Q\bar Q)
\end{eqnarray}
where the $\rho$ and $\lambda$ stand for the types of $[21]_8$ color octet configuration in Eq. (\ref{eqn::3q1}). The detailed color wave function for both color singlet and color octet states for the $q^3$ and $Q\bar Q$ are listed in Table \ref{color}.

\begin{table*}[t]
\begin{center}
\caption{\label{color}$q^3 Q\bar Q$ color wave functions.}
\begin{tabular}{p{0.15\textwidth}p{0.32\textwidth}p{0.2\textwidth}p{0.3\textwidth}}
\hline
\hline
 color list  &   $q^{3}$ color WF $\rho$ type & $q\bar q$ & $q^{3}$ color WF $\lambda$ type\\
\hline
 color singlet  & $\frac{1}{\sqrt{6}}(RGB-GRB+GBR-BGR+BRG-RBG)$ & $\frac{1}{\sqrt{3}}(R\bar R+ G\bar G + B\bar B)$  &  -  \\
\hline
 color octet 1 & $\frac{1}{\sqrt{2}}(RGR-GRR)$ & $B\bar R$ & $\frac{1}{\sqrt{6}}(2RRG-RGR-GRR)$ \\
 color octet 2 & $\frac{1}{\sqrt{2}}(RGG-GRG)$ & $B\bar G$ & $\frac{1}{\sqrt{6}}(RGG+GRG-2GGR)$ \\
 color octet 3 & $\frac{1}{\sqrt{2}}(RBR-BRR)$ & $-G\bar R$ & $\frac{1}{\sqrt{6}}(2RRB-RBR-BRR)$ \\
 color octet 4 & $\frac{1}{2}(RBG+GBR-BRG-BGR)$ & $\frac{1}{\sqrt{2}}(R\bar R- G\bar G)$ & $\frac{1}{\sqrt{12}}(2RGB+2GRB-GBR-RBG-BRG-BGR)$ \\
 color octet 5 & $\frac{1}{\sqrt{2}}(GBG-BGG)$ & $R\bar G$ & $\frac{1}{\sqrt{6}}(2GGB-GBG-BGG)$ \\
 color octet 6 & $\frac{1}{\sqrt{12}}(2RGB-2GRB-GBR+BGR-BRG+RBG)$  & $\frac{1}{\sqrt{6}}(2B\bar B- R\bar R - G\bar G)$ & $\frac{1}{2}(RBG+BRG-BGR-GBR)$ \\
 color octet 7 & $\frac{1}{\sqrt{2}}(RBB-BRB)$ & $-G\bar B$ & $\frac{1}{\sqrt{6}}(RBB+BRB-2BBR)$ \\
 color octet 8 & $\frac{1}{\sqrt{2}}(GBB-BGB)$ & $R\bar B$ & $\frac{1}{\sqrt{6}}(GBB+BGB-2BBG)$ \\
\hline
\hline
\end{tabular}
\end{center}
\end{table*}

We construct the spatial wave functions of $q^3Q\bar Q$ systems in the harmonic oscillator potential for the quark-quark interaction. A new set of relative Jacobi coordinates was introduced for the $q^3Q\bar Q$ system, different from the ones in our previous work \cite{Kai2019PRC} for $q^4\bar q$ system, the Hamiltonian for the harmonic oscillator potential is written as

\begin{gather}
H_{q^3Q^2}=\frac{\vec{p_\lambda}^2}{2m}+\frac{\vec{p_\rho}^2}{2m}+\frac{\vec{p_\sigma}^2}{2M}+\frac{\vec{p_\chi}^2}{2u_\chi}+5C(\vec{\lambda}^2+\vec{\rho}^2+\vec{\sigma}^2+\vec{\chi}^2)
\end{gather}
with $u_\chi$ being the reduced quark mass of the fourth Jacobi coordinate, defined as $u_\chi=\frac{5mM}{3m+2M}$, m and M are the mass of light quark and heavy quark respectively. $C$ is the coupling constant, and the relative Jacobi coordinates and the corresponding momenta are defined respectively as
\begin{gather*}
\vec \rho=\frac{1}{\sqrt{2}}(\vec r_1-\vec r_2)\\
\vec \lambda=\frac{1}{\sqrt{6}}(\vec r_1+\vec r_2-2\vec r_3)\\
\vec \sigma=\frac{1}{\sqrt{2}}(\vec r_4-\vec r_5)\\
\vec \chi=\frac{1}{\sqrt{30}}(2(\vec r_1+\vec r_2+\vec r_3)-3(\vec r_4+\vec r_5))\\
\vec{p_\rho}=\frac{1}{\sqrt{2}}(\vec p_1-\vec p_2)\\
\vec{p_\lambda}=\frac{1}{\sqrt{6}}(\vec p_1+\vec p_2-2\vec p_3)\\
\vec{p_\sigma}=\frac{1}{\sqrt{2}}(\vec p_4-\vec p_5)\\\vec{p_\chi}=\frac{\sqrt{5}}{\sqrt{6}}(\frac{2M(\vec p_1+\vec p_2+\vec p_3)-3m(\vec p_4+\vec p_5)}{3m+2M})
\end{gather*}
where $\vec p_i$ and $\vec r_i$ are the momenta and coordinate of $i$th quark, the antiquark is assigned the coordinate $\vec r_5$, the fourth and fifth quark form the third Jacobi coordinate $\sigma$ and the centers of first three quarks and the last two heavy quarks form the fourth Jacobi coordinate $\chi$. The permutation symmetry of pentaquarks is simply represented by the $q^3$ cluster since the $\psi_{n_\sigma,l_\sigma}(\vec\sigma)$ and $\psi_{n_\chi,l_\chi}(\vec\chi)$ is fully symmetric for any permutation between quarks. The total spatial wave function of pentaquarks may take the form,

\begin{table*}[t!]
\caption{\label{three1}Normalized $q^3$ spatial wave functions with quantum number, $N'=2n$ and $L'=M'=0$.}
\begin{ruledtabular}
\begin{tabular}{@{}ll}
$000_{[3]_S}$ & ($0,0,0,0$) \\ [2pt]
$200_{[3]_S}$ & $\frac{1}{\sqrt{2}}$($1,0,0, 0$), $\frac{1}{\sqrt{2}}$($0,0,1, 0$)  \\ [2pt]
$400_{[3]_S}$ & $\frac{\sqrt{5}}{4}$($2,0,0, 0$), $\sqrt{\frac{3}{8}}$($1,0,1, 0$), $\frac{\sqrt{5}}{4}$($0,0,2, 0$) \\[2pt]
$600_{[3]_S}$ & $\frac{\sqrt{14}}{8}$($3,0,0, 0$), $\frac{\sqrt{18}}{8}$($2,0,1, 0$), $\frac{\sqrt{18}}{8}$($1,0,2, 0$), $\frac{\sqrt{14}}{8}$($0,0,3, 0$)  \\ [2pt]
$800_{[3]_S}$ & $\frac{\sqrt{42}}{16}$($4,0,0, 0$), $\frac{\sqrt{14}}{8}$($3,0, 1, 0$), $\frac{\sqrt{15}}{8}$($2,0, 2, 0$) , $\frac{\sqrt{14}}{8}$($1,0, 3, 0$), $\frac{\sqrt{42}}{16}$($0,0,4, 0$)  \\ [2pt]
$1000_{[3]_S}$ & $\frac{\sqrt{33}}{16}$($5,0,0, 0$), $\frac{\sqrt{45}}{16}$($4,0, 1, 0$), $\frac{\sqrt{50}}{16}$($3,0, 2, 0$), $\frac{\sqrt{50}}{16}$($2,0, 3, 0$), $\frac{\sqrt{45}}{16}$($1,0, 4, 0$), $\frac{\sqrt{33}}{16}$($0,0, 5, 0$) \\ [2pt]
$1200_{[3]_S}$ & $\frac{\sqrt{429}}{64}$($6,0,0, 0$), $\frac{\sqrt{594}}{64}$($5,0, 1, 0$), $\frac{\sqrt{675}}{64}$($4,0, 2, 0$), $\frac{\sqrt{175}}{32}$($3,0, 3, 0$), $\frac{\sqrt{675}}{64}$($2,0, 4, 0$), $\frac{\sqrt{594}}{64}$($1,0, 5, 0$), \\
\phantom{} & $\frac{\sqrt{429}}{64}$($0,0,6,0$)\\[2pt]
$1400_{[3]_S}$ & $\frac{\sqrt{1430}}{128}$($7,0,0, 0$), $\frac{\sqrt{2002}}{128}$($6,0, 1, 0$), $\frac{\sqrt{2310}}{128}$($5,0, 2, 0$), $\frac{\sqrt{2450}}{128}$($4,0, 3, 0$), $\frac{\sqrt{2450}}{128}$($3,0, 4, 0$), $\frac{\sqrt{2310}}{128}$($2,0,5,0$), \\
\phantom{} & $\frac{\sqrt{2002}}{128}$($1,0, 6, 0$), $\frac{\sqrt{1430}}{128}$($0,0,7,0$) \\ [2pt]
$1600_{[3]_S}$ & $\frac{\sqrt{4862}}{256}$($8,0,0, 0$), $\frac{\sqrt{429}}{64}$($7,0, 1, 0$), $\frac{\sqrt{2002}}{128}$($6,0, 2, 0$), $\frac{\sqrt{539}}{64}$($5,0, 3, 0$), $\frac{\sqrt{2205}}{128}$($4,0, 4, 0$), $\frac{\sqrt{539}}{64}$($3,0, 5, 0$), \\
\phantom{} & $\frac{\sqrt{2002}}{128}$($2,0, 6, 0$), $\frac{\sqrt{429}}{64}$($1,0, 7, 0$), $\frac{\sqrt{4862}}{256}$($0,0,8, 0$) \\[2pt]
$1800_{[3]_S}$ & $\frac{\sqrt{4199}}{256}$($9,0,0, 0$), $\frac{\sqrt{5967}}{256}$($8,0,1, 0$), $\frac{\sqrt{1755}}{128}$($7,0, 2, 0$), $\frac{\sqrt{1911}}{128}$($6,0, 3, 0$), $\frac{\sqrt{7938}}{256}$($5,0, 4, 0$), $\frac{\sqrt{7938}}{256}$($4,0, 5, 0$), \\
\phantom{} &  $\frac{\sqrt{1911}}{128}$($3,0, 6, 0$), $\frac{\sqrt{1755}}{128}$($2,0, 7, 0$), $\frac{\sqrt{5967}}{256}$($1,0,8, 0$), $\frac{\sqrt{4199}}{256}$($0,0, 9, 0$)  \\ [2pt]
$2000_{[3]_S}$ & $\frac{\sqrt{58786}}{1024}$($10,0,0, 0$), $\frac{\sqrt{20995}}{512}$($9,0, 1, 0$), $\frac{\sqrt{99450}}{1024}$($8,0, 2, 0$), $\frac{\sqrt{6825}}{256}$($7,0, 3, 0$), $\frac{\sqrt{28665}}{512}$($6,0, 4, 0$), $\frac{\sqrt{29106}}{512}$($5,0, 5, 0$), \\
\phantom{} &  $\frac{\sqrt{28665}}{512}$($4,0, 6, 0$), $\frac{\sqrt{6825}}{256}$($3,0, 7, 0$), $\frac{\sqrt{99450}}{1024}$($2,0,8, 0$), $\frac{\sqrt{20995}}{512}$($1,0, 9, 0$), $\frac{\sqrt{58786}}{1024}$($0,0,10, 0$)  \\[2pt]
$2200_{[3]_S}$ & $\frac{\sqrt{52003}}{1024}$($11,0,0, 0$), $\frac{\sqrt{74613}}{1024}$($10,0,1, 0$), $\frac{\sqrt{88825}}{1024}$($9,0, 2, 0$), $\frac{\sqrt{98175}}{1024}$($8,0, 3, 0$), $\frac{\sqrt{103950}}{1024}$($7,0, 4, 0$), $\frac{\sqrt{106722}}{1024}$($6,0, 5, 0$), \\
\phantom{} &  $\frac{\sqrt{106722}}{1024}$($5,0,6,0$), $\frac{\sqrt{103950}}{1024}$($4,0, 7, 0$), $\frac{\sqrt{98175}}{1024}$($3,0,8, 0$), $\frac{\sqrt{88825}}{1024}$($2,0, 9, 0$), $\frac{\sqrt{74613}}{1024}$($1,0,10, 0$), $\frac{\sqrt{52003}}{1024}$($0,0,11, 0$) \\ [2pt]
\end{tabular}
\end{ruledtabular}
\end{table*}

\begin{table*}[t]
\caption{\label{norp4}$q^3 Q\bar Q$ pentaquark spatial wave functions of symmetric type.}
\begin{ruledtabular}
\begin{tabular}{@{}ll}
     $\phantom{-}\Psi^{q^{3} Q\bar Q}_{000_{[5]_S}}$ & $\psi^{q^3}_{000_{[3]_S}}\psi_{0,0}(\vec\sigma\,)\psi_{0,0}(\vec\chi\,)$\\[2pt]
   $\phantom{-}\Psi^{q^{3} Q\bar Q}_{200_{[5]_S}}$ & $\psi^{q^3}_{200_{[3]_S}}\psi_{0,0}(\vec\sigma\,)\psi_{0,0}(\vec\chi\,)$, $\psi^{q^3}_{000_{[3]_S}}\psi_{1,0}(\vec\sigma\,)\psi_{0,0}(\vec\chi\,)$, $\psi^{q^3}_{000_{[3]_S}}\psi_{0,0}(\vec\sigma\,)\psi_{1,0}(\vec\chi\,)$\\[2pt]
   $\phantom{-}\Psi^{q^{3} Q\bar Q}_{400_{[5]_S}}$ & $\psi^{q^3}_{400_{[3]_S}}\psi_{0,0}(\vec\sigma\,)\psi_{0,0}(\vec\chi\,)$, $\psi^{q^3}_{200_{[3]_S}}\psi_{1,0}(\vec\sigma\,)\psi_{0,0}(\vec\chi\,)$, $\psi^{q^3}_{200_{[3]_S}}\psi_{0,0}(\vec\sigma\,)\psi_{1,0}(\vec\chi\,)$, $\psi^{q^3}_{000_{[3]_S}}\psi_{2,0}(\vec\sigma\,)\psi_{0,0}(\vec\chi\,)$, \\
   $\phantom{-}$ & $\psi^{q^3}_{000_{[3]_S}}\psi_{1,0}(\vec\sigma\,)\psi_{1,0}(\vec\chi\,)$, $\psi^{q^3}_{000_{[3]_S}}\psi_{0,0}(\vec\sigma\,)\psi_{1,0}(\vec\chi\,)$ \\[2pt]
   $\phantom{-}\Psi^{q^{3} Q\bar Q}_{600_{[5]_S}}$ & $\psi^{q^3}_{600_{[3]_S}}\psi_{0,0}(\vec\sigma\,)\psi_{0,0}(\vec\chi\,)$, $\psi^{q^3}_{400_{[3]_S}}\psi_{1,0}(\vec\sigma\,)\psi_{0,0}(\vec\chi\,)$, $\psi^{q^3}_{400_{[3]_S}}\psi_{0,0}(\vec\sigma\,)\psi_{1,0}(\vec\chi\,)$, $\psi^{q^3}_{200_{[3]_S}}\psi_{2,0}(\vec\sigma\,)\psi_{0,0}(\vec\chi\,)$,  \\
   $\phantom{-}$ & $\psi^{q^3}_{200_{[3]_S}}\psi_{1,0}(\vec\sigma\,)\psi_{1,0}(\vec\chi\,)$, $\psi^{q^3}_{200_{[3]_S}}\psi_{0,0}(\vec\sigma\,)\psi_{2,0}(\vec\chi\,)$, $\psi^{q^3}_{000_{[3]_S}}\psi_{3,0}(\vec\sigma\,)\psi_{0,0}(\vec\chi\,)$, $\psi^{q^3}_{000_{[3]_S}}\psi_{2,0}(\vec\sigma\,)\psi_{1,0}(\vec\chi\,)$, \\
      $\phantom{-}$ & $\psi^{q^3}_{000_{[3]_S}}\psi_{1,0}(\vec\sigma\,)\psi_{2,0}(\vec\chi\,)$, $\psi^{q^3}_{000_{[3]_S}}\psi_{0,0}(\vec\sigma\,)\psi_{3,0}(\vec\chi\,)$\\[2pt]
   $\phantom{-}\Psi^{q^{3} Q\bar Q}_{800_{[5]_S}}$ & $\psi^{q^3}_{800_{[3]_S}}\psi_{0,0}(\vec\sigma\,)\psi_{0,0}(\vec\chi\,)$, $\psi^{q^3}_{600_{[3]_S}}\psi_{1,0}(\vec\sigma\,)\psi_{0,0}(\vec\chi\,)$, $\psi^{q^3}_{600_{[3]_S}}\psi_{0,0}(\vec\sigma\,)\psi_{1,0}(\vec\chi\,)$, $\psi^{q^3}_{400_{[3]_S}}\psi_{2,0}(\vec\sigma\,)\psi_{0,0}(\vec\chi\,)$, \\
\phantom{-} & $\psi^{q^3}_{400_{[3]_S}}\psi_{1,0}(\vec\sigma\,)\psi_{1,0}(\vec\chi\,)$, $\psi^{q^3}_{400_{[3]_S}}\psi_{0,0}(\vec\sigma\,)\psi_{2,0}(\vec\chi\,)$, $\psi^{q^3}_{200_{[3]_S}}\psi_{3,0}(\vec\sigma\,)\psi_{0,0}(\vec\chi\,)$, $\psi^{q^3}_{200_{[3]_S}}\psi_{2,0}(\vec\sigma\,)\psi_{1,0}(\vec\chi\,)$, \\
\phantom{-} & $\psi^{q^3}_{200_{[3]_S}}\psi_{1,0}(\vec\sigma\,)\psi_{2,0}(\vec\chi\,)$, $\psi^{q^3}_{200_{[3]_S}}\psi_{0,0}(\vec\sigma\,)\psi_{3,0}(\vec\chi\,)$, $\psi^{q^3}_{000_{[3]_S}}\psi_{4,0}(\vec\sigma\,)\psi_{0,0}(\vec\chi\,)$, $\psi^{q^3}_{000_{[3]_S}}\psi_{3,0}(\vec\sigma\,)\psi_{1,0}(\vec\chi\,)$,\\
\phantom{-} & $\psi^{q^3}_{000_{[3]_S}}\psi_{2,0}(\vec\sigma\,)\psi_{2,0}(\vec\chi\,)$, $\psi^{q^3}_{000_{[3]_S}}\psi_{1,0}(\vec\sigma\,)\psi_{3,0}(\vec\chi\,)$, $\psi^{q^3}_{000_{[3]_S}}\psi_{0,0}(\vec\sigma\,)\psi_{4,0}(\vec\chi\,)$ \\[2pt]
\end{tabular}
\end{ruledtabular}
\end{table*}

\begin{eqnarray}\label{eqn::swf}
 \Psi_{NLM}^{[X]_y}= \psi^{q^3[X]_y}_{N'L'M'}\otimes\psi_{n_\sigma,l_\sigma}(\vec\sigma)\otimes\psi_{n_\chi,l_\chi}(\vec\chi)
\end{eqnarray}
which is simply the product of the $q^3$ spatial wave function shown in Table \ref{three1} and the harmonic oscillator wave functions $\psi_{n_\sigma,l_\sigma}(\vec\sigma\,)$ and $\psi_{n_\chi,l_\chi}(\vec\chi\,)$ for the Jacobi coordinate $\sigma$ and $\chi$. $[X]_{y}$ stands for all possible permutation symmetries of the $q^3$ cluster, where, $[X]_{y} = \{{[3]_S},\;{[21]_{\rho,\lambda}},\;{[111]_{A}}\}$.
$N$, $L$, and $M$ are respectively the total principle quantum number, total angular momentum and magnetic quantum number of the pentaquark  ($l_\sigma=0$, $l_\chi=0$), with
\begin{eqnarray}\label{eqn::q4xi}
N &=& 2n_{\rho}+ l_{\rho}+2n_{\lambda}+l_{\lambda}+2n_{\sigma}+ l_{\sigma}+2n_{\chi}+l_{\chi}
\end{eqnarray}

The spatial wave functions of the $q^3$ subsystem of $q^3 Q \bar Q$ pentaquarks with the permutation symmetries $[3]_S$ are listed in Table \ref{three1} up to $N'=22$, where $l_\rho$, $l_\lambda$, and are $L'$ are limited to $0$, $1$ and $2$ only. To save space, we show only the symmetric spatial wave function while the spatial wave function for other possible permutation symmetries $\{{[21]_{\rho,\lambda}}$, and ${[111]_{A}}$ will not be specified here. Note that we have set $M'=0$ and used the abbreviation, 
\begin{eqnarray}\label{eqn::not3}
&& \sum_{\{n_i,l_i,m_i\}} C_{n_{\rho}, l_{\rho},m_{\rho},n_{\lambda},l_{\lambda},m_{\lambda}} \psi_{n_{\rho}l_{\rho}m_{\rho}}(\vec\rho\,)\psi_{n_{\lambda}l_{\lambda}m_{\lambda}}(\vec\lambda\,) \nonumber \\
\nonumber \\
&\equiv&  \sum_{\{n_i,l_i\}} C_{n_{\rho}, l_{\rho} ,n_{\lambda},l_{\lambda}}\,\psi(n_{\rho},l_{\rho},n_{\lambda},l_{\lambda})  \nonumber \\
& \equiv&  \sum_{\{n_i,l_i\}} C_{n_{\rho}, l_{\rho} ,n_{\lambda},l_{\lambda}}\,(n_{\rho},l_{\rho},n_{\lambda},l_{\lambda}) 
\end{eqnarray}

The spatial wave functions of pentaquarks with the $q^3 Q\bar Q$ symmetry $[5]_S$ are listed in the Table \ref{norp4} (Up to $N=14$ energy level is sufficient for the numerical calculations), where $\psi^{q^{3}}_{N'L'M'}$ ($L'=M'=0$) and
$\psi_{n_\sigma,l_\sigma}(\vec\sigma\,)$ ($l_\sigma=0$), $\psi_{n_\chi,l_\chi}(\vec\chi\,)$ ($l_\chi=0$) are
the spatial wave functions of the $q^3$ subsystem and the harmonic oscillator wave function for the $\vec\sigma$ and $\vec\chi$ coordinates, respectively.
Without any limitation for $n_\sigma$ and $n_\chi$, all degenerate states of each pentaquark energy level up to $N=14$ served as a complete basis. $N\leq8$ states are listed below, the higher ones follow the rule that $N=N_{q^3}+2(n_\sigma+n_\chi)$.

\bibliographystyle{unsrt}

\end{document}